\documentclass[11pt]{iopart}

\usepackage{epsf}
\usepackage{longtable}
\usepackage{hyperref}
\usepackage{graphics,epsfig,placeins,subfig,wrapfig}
\expandafter\let\csname equation*\endcsname\relax

\expandafter\let\csname endequation*\endcsname\relax
\usepackage{amsmath,amssymb,calc,amsfonts}
\usepackage[usenames]{color}
\usepackage{soul}
\usepackage{amsfonts}
\usepackage[mathscr]{euscript}

\DeclareMathAlphabet\mathbfcal{OMS}{cmsy}{b}{n}

\def\eadnew#1#2{\address{#2 E-mail: \mailto{#1}}} 

\begin{document}

\title{Shadows around the $q$-metric}

\author{J. A. Arrieta-Villamizar $^{1,*}$, 
J. M. Vel\'asquez-Cadavid $^{1,**}$,
O.  M. Pimentel $^{1,\dag}$,
F. D. Lora-Clavijo $^{1,\ddag}$,
A. C. Guti\'errez-Pi\~neres $^{1,2, \checkmark}$}
\address{$^1$  Escuela de F\'isica, Universidad Industrial de Santander, A. A. 678,
Bucaramanga 680002, Colombia.}
\address{$2$  Instituto de Ciencias Nucleares, Universidad Nacional Aut\'onoma de M\'exico,
 \\AP 70543,  M\'exico, DF 04510, M\'exico}

\eadnew{jesus2208058@correo.uis.edu.co}{$^{*}$}
\eadnew{juan2208056@correo.uis.edu.co}{$^{**}$}
\eadnew{oscar.pimentel@correo.uis.edu.co}{$^{\dag}$}
\eadnew{fadulora@uis.edu.co}{$^{\ddag}$}
\eadnew{acgutier@uis.edu.co}{$^{\checkmark}$}

\pacs{04.20.-q, 04.20.Dw, 42.15.Dp}

\begin{abstract}

One crucial problem in relativistic astrophysics is that of the nature of black hole candidates. It is usually assumed that astrophysical black holes are described by the Schwarzschild or Kerr space-times; however, there is no direct evidence to assert this. Moreover, there are various solutions in general relativity that can be alternatives to black holes, usually called black hole mimickers. In this work, we study the shadow produced by a compact object described by the $q$-metric, which is the simplest static and axially symmetric solution of Einstein equations with a non-vanishing quadrupole moment. This particular space-time has the property of containing an independent parameter $q$, which is related to the compact object deformation. The solution corresponds to naked singularities for some specific values of this parameter.
Additionally, we analyze the eigenvalues of the Riemann tensor using the $SO(3, C)$ representation, which allows us to find, in an invariant way, regions where there may be repulsive effects. Furthermore, we numerically solve the motion equations to show the shadow, the Einstein ring, and the gravitational lensing to establish a possible signature of such repulsive effects. We found that as $q$ is smaller, the Einstein ring decreases, but the shape is the same as the Schwarzschild black hole case. However, for values of $q$ lower or equal than $-0.5$, repulsive gravitational effects appear in the gravitational lensing close to the compact object, where a strong dependence of the system to the initial conditions seems to take place.
\end{abstract}

\section{Introduction}
Recently, the Event Horizon Telescope Collaboration (EHT) imaged for the first time a real supermassive black hole with a mass of $M=(6.7\pm 0.7)\times 10^9 M_{\odot}$ and a dimensionless spin parameter $a = 0.94$ \cite{MM87} at the center of the galaxy M87 \cite{EHT19}, consistent with the shadow produced by a Kerr black hole. For this reason, the EHT measurements are of great importance, since this results can be used to constrain the background metric, and are complementary with the stars trajectories measurements around compact objects given by the GRAVITY collaboration \cite{abuter2018detection}, and the LIGO/VIRGO gravitational wave detections, associated to the dynamics of the space-time \cite{abbott2016observation}. 

Even though the EHT images can rule out some black holes alternatives, there isn't definite evidence or proof that all compact astrophysical objects can be described by the Schwarzschild or Kerr black holes \cite{shaikh2018shadows, abramowicz2002no}. Besides these astrophysical relevant solutions to the general relativity, there are other solutions without an event horizon that can mimic the electromagnetic signature of a black hole \cite{lemos2008black}, making challenging the identification of a compact object from initial observations. In fact, according to \cite{shaikh2019can}, any object with a photon sphere can produce very similar shadows to that of black holes. Some examples of black hole mimickers without curvature singularities are the boson stars \cite{kaup1968klein, LoraClavijo:2010xc, Liebling2012}, whose shadows are very similar to those obtained with the Kerr solution \cite{vincent2016imaging}, and the gravastars, derived from the concept of Bose-Einstein condensation \cite{mazur2004gravitational, sakai2014gravastar}. 

Additionally, it has been shown that starting from physically reasonable conditions it is possible to collapse a fluid to form a compact object with a curvature singularity, but without an event horizon. Those systems are called naked singularities to distinguish them from black holes, whose defining feature is the presence of an event horizon \cite{shaikh2018shadows, joshi2013distinguishing, joshi2011equilibrium, joshi2011recent, singh1999gravitational, shapiro1991formation}. These results violate the unproved cosmic censorship conjecture \cite{christodoulou1984violation, shapiro1991formation}, which states that it is not possible to form a singularity without an event horizon \cite{penrose2002gravitational}, so they establish the possibility to study and observe gravitational phenomena in the ultra-strong regime, i.e., in the vicinity of a curvature singularity \cite{shaikh2018shadows}. That is why a lot of effort has been made to obtain the image of naked singularities in order to determine the differences with black holes images \cite{joshi2013distinguishing, babar2017periodic, liu2018distinguishing, gyulchev2019image}. 

Naked singularities obtained as exact solutions to the Einstein equations are important because they usually have a clear interpretation in terms of their multipole moments. The shadow of charged compact objects described by the Reissner-Nordstrom, and Kerr-Newman space-times \cite{joshi2013distinguishing} were obtained in \cite{takahashi2005black, zakharov2012shadows}. These metrics posses naked singularities for some particular values of their parameters. An interesting solution to the Einstein equations is the one obtained by A. Janis, E. Newman, and J. Winicour, which generalizes the Schwarzschild vacuum solution, including a massless scalar field \cite{janis1968reality}. A geodesic analysis of this space-time is presented in \cite{zhou2015geodesic}, and the image of its naked singularity is obtained in \cite{gyulchev2019image} when the photon source is a thin accretion disc. On the other hand, it is possible to get an exact solution for a static source with non-zero quadrupole moment by applying the Zipoy-Voorhees transformations \cite{zipoy1966topology, voorhees1970static} to the Schwarzschild metric \cite{quevedo2011mass}. The resulting space-time corresponds to the $q$-metric.

This space-time has been studied because it is the simplest static and axially symmetric solution to the Einstein equations with a non-vanishing quadrupole moment, so it is very useful to determine the observable effects associated with an independent deformation parameter $q$. Additionally, for some values of this parameter, the solution represents a naked singularity with regions of repulsive gravity \cite{quevedo2011mass}. These properties are interesting because they can be used to investigate the differences between black holes and naked singularities in a static space-time. In a previous work, one of us and their collaborators investigated the motion of a test particle in the field of the $q$-metric \cite{boshkayev2016motion}, and show that the geodesic motion drastically depends on the deformation parameter. The results are potentially applied to accretion discs, and in particular, they are useful to compute the deformation parameter from the inner radius of the disc. One description of the Zipoy-Vorhees generalization of the Schwarzschild solution is known as the $\gamma$-metric, where $\gamma = 1 + q$. The shadow associated with this metric was computed in \cite{abdikamalov2019black}, where the authors analyzed the image cast by oblate and prolate sources and studied the gravitational lensing in the equatorial plane by measuring the deflection angle for photons. They concluded that for some values of the deformation parameter, the naked singularity mimics black holes, while for other values, the results differ significantly from the Schwarzschild case and can be explained in terms of a repulsive gravitational field effect near the singular surface.

Since repulsive gravity is a remarkable property of naked singularities, in this paper, we apply the method proposed by O. Luongo and H. Quevedo \cite{luongo2014characterizing} to the naked singularities described by the $q$-metric, to define invariantly the regions where the repulsive gravitational field has effects. Then we use our ray-tracing code to obtain and analyze the image and gravitational lensing produced by photons for different values of the $q$ parameter and determine an observable region where the repulsive gravity effects take place. It is worth mentioning that unlike in the case of the $\gamma$-metric \cite{abdikamalov2019black}, in this work, we image the whole gravitational lensing produced by the $q$-metric as well as the Einstein ring as a function of the deformation parameter $q$. This paper is organized as follows. In Sec. \ref{sec:ST}, we describe briefly the $q$-metric and as well as the invariant eigenvalues of Riemann curvature tensor, using the $SO(3, C)$ representation. In Sec. \ref{sec:PME}, we show the photon motion equations, from which it is possible to simulate the image cast by a compact object through the shadow formulation and the backward ray-tracing method. In Sec. \ref{sec:QM}, we present the results obtained by our numerical simulations of the shadow generated by a compact object described by the $q$-metric for different values of the mass deformation parameter $q$. Finally, in Sec. \ref{sec:conclusions} we present the conclusions of our work. Along this paper we employ the signature $(-,+,+,+)$ and geometrized units where $G=c=1$. 

\section{$q$-metric spacetime}
\label{sec:ST}

The effect of repulsive gravity in the dynamics of massive particles is getting more attention in the strong gravitational field community, so an invariant definition of repulsion radius has been proposed. Similarly, discussions on issues that seem to have not counterpart in the Newtonian theory of gravitation stay currently open. For example, one of the most important open questions in general relativity is related to the Israel-Carter conjecture, informally known as ``black holes have no hair."  According to this theorem, the only information no radiated away during the collapse process and, consequently, the unique necessary knowledge required to describe a Kerr black hole comes from its mass and angular momentum.

Likewise, the Penrose cosmic censorship conjecture states that singularities produced by the gravitational collapse of nonsingular, asymptotically flat initial data, will always be surrounded by event horizons and hence can never be visible from the outside (no naked singularities) \cite{PhysRevLett.66.994}. Nevertheless, currently, both Israel-Carter conjecture and Penrose cosmic censorship conjecture lack conclusive evidence, so they correspond only to a hypothesis. Conclusions from several tests seem to indicate that the final state of continuous collapse can turn out to be either a black hole or a naked singularity \cite{joshi2007gravitational}. Therefore, many questions about the final state of gravitationally collapsing objects arise without conclusive answers. At present, much attention is devoted to the discussion on distinguishing black holes from naked singularities, both from a theoretical and observational point of view. Increasing efforts are today dedicated to examining shadows of black holes and naked singularities.  Shaikh et al. \cite{shaikh2018shadows}  conclude that while black holes imply shadows, the converse is not valid and that certain naked singularities could produce a shadow as well. 

The Schwarzschild solution in Weyl coordinates admits the interpretation of the Newtonian potential for a rod with deformation parameter $\delta \equiv \mu/\lambda =1$, where $\mu$ and $\lambda$ correspond to the mass and length of the rod, respectively. Bach and Weyl \cite{weyl1922nene} and Darmois \cite{darmois1927memorial} discovered a family of solutions for arbitrary values of $\delta$, which generalizes the Schwarzschild solution.  It is now commonly referred to either as the Zipoy-Vorhees solution or as the $\gamma$-metric \cite{griffiths2009exact}. Properties of these solutions have been studied by Zipoy \cite{zipoy1966topology}, Gautreau and Anderson \cite{gautreau1967directional}, Bonnor and Sackfield \cite{bonnor1968interpretation}, Vorhees \cite{voorhees1970static},
 Esposito and Witten \cite{esposito1975static}, and many others. Kodama and Hikida \cite{kodama2003global} have shown that this line element is geometrically point-like for  $ \delta < 0$, rod-like for $0 < \delta < 1$, and ring-like for $\delta >1$ . In all of the cases, the singularities are always naked.

Quevedo \cite{quevedo2011mass} studied this family by considering the deformation through $\delta = 1 + q$ for arbitrary values of $q$. This metric describes the exterior field of static deformed mass distribution. The lowest mass and quadrupole moments are $M_{0}=(1 + q)m$ and $M_{2} = -m^3q(1+q)(2+q)/3$, respectively. Because the deformation is described by the quadrupole moment of the mass distribution since $M_{0}$ and $M_{2}$ only depends on the values of $m$ and $q$, the parameter $q$ determines its deformation uniquely. It is worth mentioning that the mass moment $M_0$, which corresponds to AMD mass, becomes zero when $q=-1$, so the Minkowski space-time is recovered as it is expected.

Direct inspection of the Kestchman scalar shows that the solution exhibits a singularity at the origin of coordinates and at the hypersurface $r=2m$. Both are present for any values of $q$ except for $q = 0$, where the Schwarzschild space-time is recovered. Additionally, a curvature singularity occurs at the surface given by the equation $$ r^2 - 2mr + m^2\sin^2\theta =0,$$ for values of $q$ in the interval $(-1, -1+\sqrt{3/2})\setminus\{0\}$. These singularities correspond to naked singularities without event horizons and as pointed out by Luongo and Quevedo \cite{luongo2014characterizing}
surrounded by possible regions of repulsive gravity.

 No matter how small the naked singularity's quadrupole is, it always affects test particles' motion \cite{boshkayev2016motion}.
 If we consider unbounded orbits with specific non zero initial radial and angular velocities, all the test particles can escape from the naked singularity's gravitational field.  
The quadrupole's small values  (e.g., $q = \pm 0.099$ ) determine only the direction along which the particle escapes towards infinity. 
Similarly, for small values (e.g., $q = 0.009$ ), stable circular geodesic of a particle with zero initial radial velocity moves slowly towards the central singularity.  Finally, it reaches a stable circular orbit with a radius smaller than the initial radius.  
We see that the quadrupole with $q=\pm 0.022 $ does not affect the geodesic corresponding orbit's bounded character with zero initial radial velocity and nonzero value for the initial angular velocity. However, it can drastically modify the geometric structure of the trajectory.

We have two reasons to do not use the $\gamma$-metric interpretation of the Zipoy-Vorhees generalization of the Schwarzschild solution in favor of the $q$-metric interpretation. First, because it provides knowledge about the multipole structure of the sources and the asymptotic behavior of the space-time. Second, because that information could indicate the places in the space-time for the possible regions of repulsive gravity. 

In this section, we investigate this space-time, and use a method to characterize the possible regions with repulsive gravity in an invariant way, which is defined through the eigenvalues of the Riemann tensor. The idea is quite simple: since the curvature tensor is a measure of the gravitational interaction, the curvature eigenvalues with their scalar property provide an invariant measure of the gravitational interaction \cite{gutierrez2019c}. It is possible to represent the curvature tensor as a (6$\times$6)-matrix by introducing the bivector indices, which encode the information of two different tetrad indices. Then, the Riemann tensor can be represented by a symmetric matrix with 20 independent components. This correspondence can be applied to all the irreducible components of the Riemann tensor. So, the curvature tensor can be expressed as a (3$\times$3)-matrix that represents the irreducible pieces of the curvature with respect to the Lorentz group $SO(3, 1)$, that is isomorphic to the group $SO(3, C)$. Thus, it is possible to introduce a  complex local basis  where the curvature is given as a (3$\times$3)-matrix. This is the so-called $SO(3, C)$-representation of the Riemann tensor  \cite{gutierrez2019c, synge1964petrov, debever1964rayonnement}.

We use this representation to obtain the eigenvalues of the curvature tensor.  In the following lines,  we explain some details concerning to  its implementation. 
The Riemann curvature tensor locally measured the components that can be decomposed into a sum of (6$\times$6)-matrices, which are irreducible representations of the full Lorentz group, 
\begin{align*}
\mathbf{\cal R} =  \mathbf{\cal W} + \mathbf{\cal E} + \mathbf{\cal S}, 
\end{align*}
being
\begin{align*}
\mathbf{\cal W} =
  \left[ {\begin{array}{cc}
   M  &    N\\
   N  &  - M  
  \end{array} } \right]  \ , \quad
  \mathbf{\cal E} =
  \left[ {\begin{array}{cc}
   P  &    Q\\
   Q  &  - P 
  \end{array} } \right] \ , \quad
  \mathbf{\cal S} = \frac{R}{12}
  \left[ {\begin{array}{cc}
   -I   &    0\\
    0  &  I  
  \end{array} } \right]  \ ,
  \end{align*}
where $M, N, P $ are real symmetric (3$\times$3)-matrices and $Q$ is real and antisymmetric matrix. Finally, $R$  and $I$ denote the Ricci curvature scalar and  the (3$\times$3)-unit matrix, respectively. 
  The matrices $M$ and $N$ are trace-free.
Hence, according to this decomposition, the Riemann tensor's symmetries make it possible to write the eigenvalues equation
\begin{align}\label{eq:eingevalues}
\mathbf{\cal K}  \mathbf{\cal Z} = \Lambda \mathbf{\cal Z} \ , \qquad \det |\mathbf{\cal K} - \Lambda I | =0 , 
\end{align}
in which
$$
\mathbf{\cal K} =  M + P - \frac{R}{12} I -  i  ( Q  +  N)   \quad \text{and} \quad     \mathbf{\cal Z}  = X  +  i \  Y .$$
Eigenbivectores  $\mathbf{\cal Z}$  are defined through  anstisymmetric  tensors of second order $X$, $Y$, which represent a 6-dimensional vector space.  

There are six equations in the eigenvalues equation for the (6$\times$6)-representation of the Riemann curvature tensor, and the consistency condition yields six eigenvalues.  
They do not need to be all distinct, if complex occurs in conjugate pairs. As it can be seen from equation (\ref{eq:eingevalues}), there are three equations, whose consistency condition is given by the corresponding (3$\times$3) determinant, which yield the eigenvalues $\Lambda$. Therefore, we achieve the curvature tensor's eigenvalues in its $SO(3, C)$-representation by attempting these conditions.
Finally, note that three roots of the eigenvalues equation are the three complex conjugates of (\ref{eq:eingevalues}).
 
In spherical coordinates, the $q$-metric can be written by the metric tensor 
  \begin{align}\label{eq:q-metric}
  \mathbf{g} &= \left(1 - \frac{2m}{r}\right)^{1 + q}\operatorname{d}{t} \otimes \operatorname{d}{t} 
            - \left(1 - \frac{2m}{r}\right)^{ - q}  \Bigg[  \left( 1 + \frac{m^2\sin^2{\theta}}{r^2 - 2mr}  \right)^{-q(2 +q)} \\
     & \times \Bigg( \operatorname{d}{r} \otimes \operatorname{d}{r}\Bigg(1 - \frac{2m}{r}\Bigg)^{-1}
      \nonumber + r^2 \operatorname{d}{\theta} \otimes \operatorname{d}{\theta} \Bigg) 
      + r^2 \sin^2 {\theta} \operatorname{d}{\varphi} \otimes \operatorname{d}{\varphi} \Bigg].
   \end{align}   

So, by using the $SO(3, C)$-representation the eigenvalues of Riemann curvature tensor, corresponding to the $q$-metric, can be written as
\begin{gather*}
 \Lambda_{I} = \frac{ m(1 + q)}{r^3} \left( 1 - \frac{2 m}{ r} \right)^{-q^2 - q -1} \left( \frac{m^2\sin^2\theta}{r^2} + 1 - \frac{2M}{r} \right)^{q(2+q)} %
\times \left( 1 - \frac{(2+q)M}{r} \right) , \\ \\
 \Lambda_{II} = - \frac{m(1+q)}{2r^3} \left( 1 - \frac{2m}{r}\right)^{-q^2 - q-1} \left( \frac{m^2\sin^2\theta}{r^2} + 1 - \frac{2m}{r} \right)^{q(2+q)} \\ 
\times \left[ \left( 1 - \frac{(2+q)m}{r}\right) - \frac{\sqrt{D}}{r^2} \left( \frac{m^2\sin^2\theta}{r^2} + 1 - \frac{2m}{r} \right)^{-\frac{1}{2}} \right], \\ \\
 \Lambda_{III} = - \frac{m(1+q)}{2r^3} \left( 1 - \frac{2m}{r}\right)^{-q^2 - q-1} \left( \frac{M^2\sin^2\theta}{r^2} + 1 - \frac{2m}{r} \right)^{q(2+q)} \\ 
\times \left[ \left( 1 - \frac{(2+q)m}{r}\right) +\frac{\sqrt{D}}{r^2} \left( \frac{m^2\sin^2\theta}{r^2} + 1 - \frac{2m}{r} \right)^{-\frac{1}{2}} \right],
\end{gather*}
where
\begin{align*}
D & \equiv \bigg( 3r^2 - 3(3+ q)mr + (6 +7q + 2q^2 )m^2 \bigg)^2 \\
  & - m^2\cos^2\theta \left(3(1+ 2q )(3+ 2q )r^2 -6 (3+ 6q + 2q^2)(2+q)Mr + (2 +q)^2(3 +2q )^2m^2 \right).
\end{align*}
Notice that $\Lambda_{I}+\Lambda_{II}+\Lambda_{III} =0$, because the $q$-metric is a solution of the Einstein field equations \cite{gutierrez2019c}
 and in the limiting case of a vanishing deformation parameter ($q = 0$), we obtain
 $$\Lambda_{I}=\frac{m}{r^3}, \ \Lambda_{II}=-\frac{2m}{r^3},\ \Lambda_{III}=\frac{m}{r^3},$$ 
 which are the eigenvalues of the Riemann tensor for the Schwarzschild solution.
 
\subsection{Repulsive gravity around a naked singularity}
For the sake of completeness, we will briefly recall the conjecture established in \cite{luongo2014characterizing, gutierrez2019c}, which asserts that one can use the eigenvalues of the Riemann tensor to detect the regions of the gravitational field of compact objects where repulsive effects are of importance. We start by indicating that since the gravitational field of compact objects is asymptotically flat, the eigenvalues should vanish at infinity. As we approach the central source, the eigenvalues will increase until they become infinity at the singularity if gravity is everywhere attractive or exchange their sign at some point, indicating the character of gravity has changed. At the position where the eigenvalue vanishes, the attractive gravity becomes entirely compensated by the action of repulsive gravity at that point. Next, we perform a formal formulation of the above reasoning. As conjectured above, the presence of an extremum in an eigenvalue indicates the existence of repulsive gravity. Let $\{\Lambda_i \}$, $i=1,2,3$ represents the set of eigenvalues of an exterior space-time. Then, let $\{ r_l \}$, $l=1,2, \dots$ with $0 < r_l < \infty$ represents the set of solutions of the equation 
$$\frac{\partial \Lambda_i}{\partial r}\Big{|}_{r=r_l} =0, 
\qquad
\text{with}
\qquad
r_{\text{rep}} = \text{max}\{r_l \},$$
i.e. $r_{\text{rep}}$ is the largest extremum of the eigenvalues and is called repulsion radius. We will show that in the $q$-metric space-time, there are several extrema, and consequently, possible regions of repulsive gravity.

\begin{figure}
\centering
\subfloat[$q= -0.30$]{
 \includegraphics[width = 0.5\textwidth]{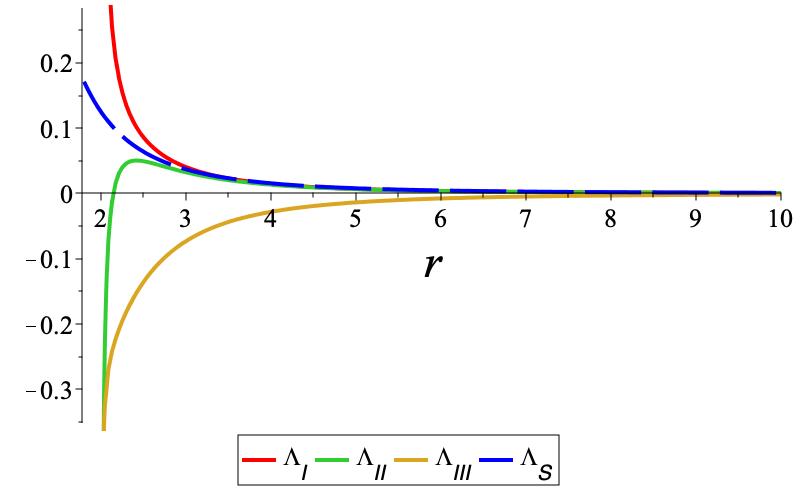}
}
\subfloat[$q= -0.50$]{
 \includegraphics[width = 0.5\textwidth]{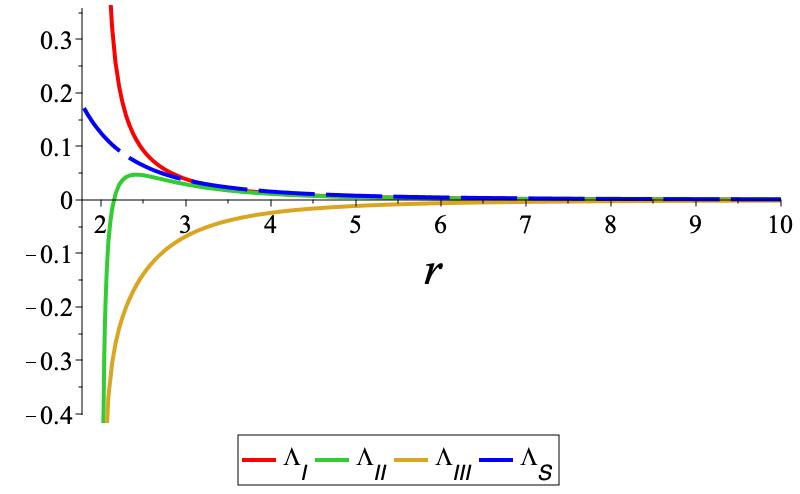}
}
\hspace{0mm}
\subfloat[$q= -0.64$]{
 \includegraphics[width = 0.5\textwidth]{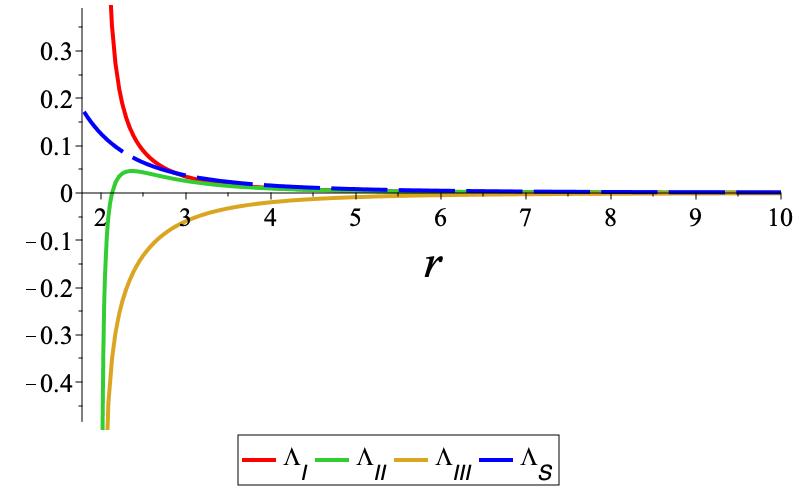}
}
\subfloat[$q= -0.74$]{
 \includegraphics[width = 0.5\textwidth]{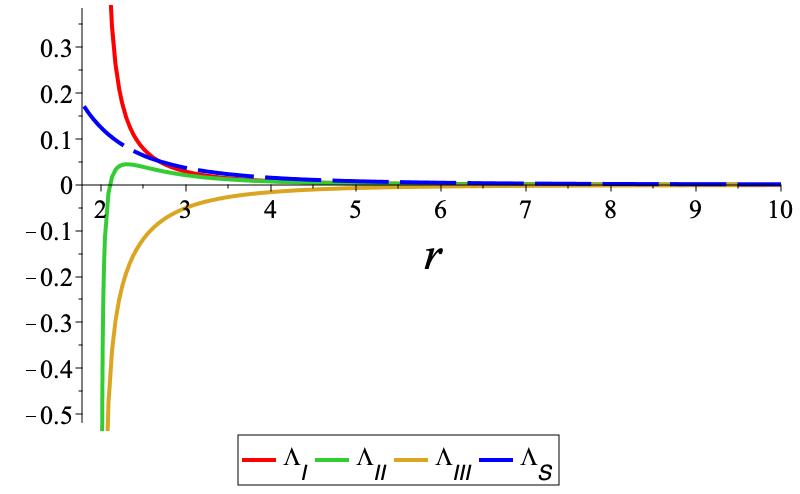}
}
\caption{ Eigenvalues of the Riemann tensor of the $q$-metric in the plane $\theta = \pi/2$ plotted for different values of the deformation parameter $q$. The dashed blue line in each panel represents the second eigenvalue of the Riemann tensor of a Schwarzschild space-time (limiting case $q = 0$). As we can see, the second eigenvalues $\Lambda_{II}$ shows the presence of a zone with repulsive gravity.}
\label{fig:ThreeEigenvaluesA}
\end{figure}

\begin{figure}[h!]
\centering
\subfloat[$q= -0.30$]{
  \includegraphics[width = 0.5\textwidth]{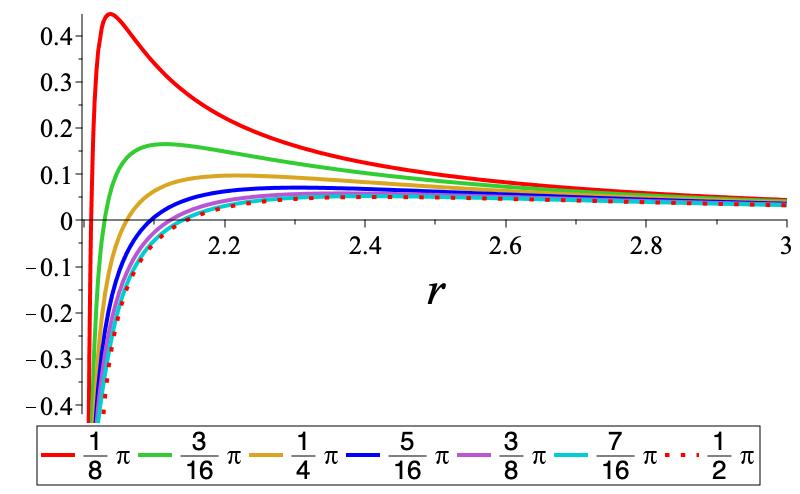}
}
\subfloat[$q= -0.50$]{
  \includegraphics[width = 0.5\textwidth]{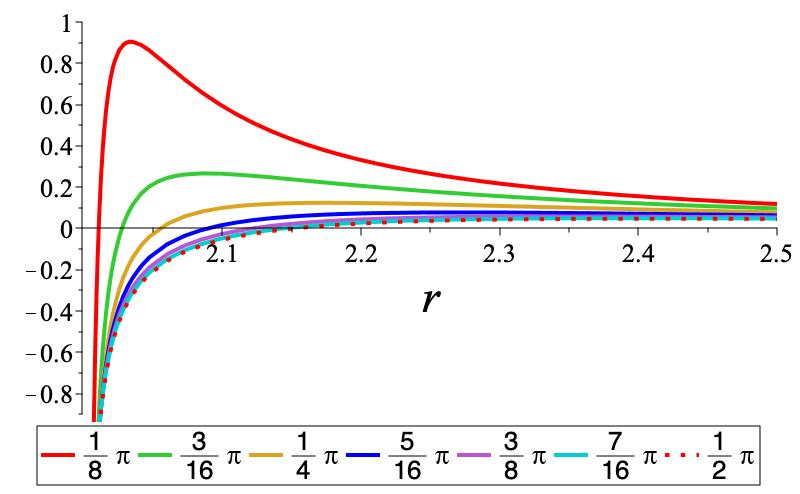}
}
\hspace{0mm}
\subfloat[$q= -0.64$]{
  \includegraphics[width = 0.5\textwidth]{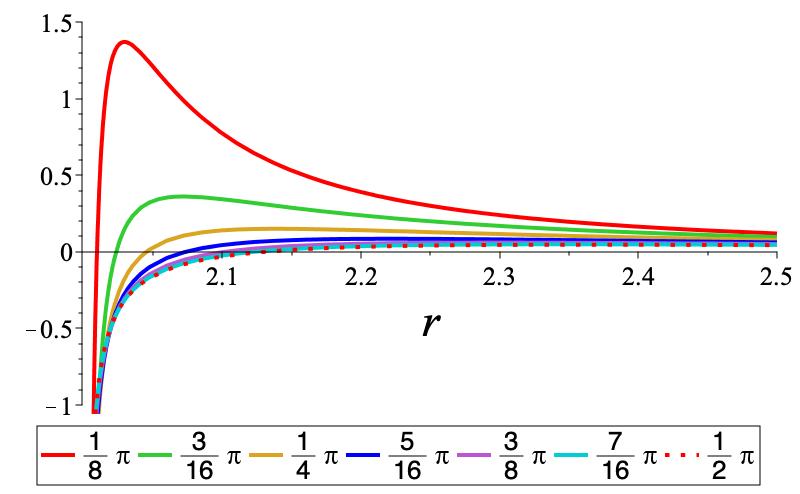}
}
\subfloat[$q= -0.74$]{
  \includegraphics[width = 0.5\textwidth]{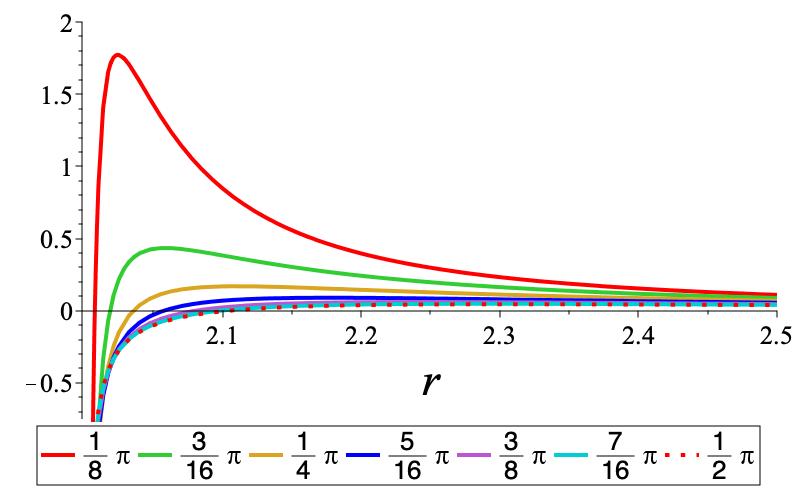}
}
\caption{ The second eigenvalue of the Riemann tensor $\Lambda_{II}$ corresponding to the $q$-metric in terms of the radial distance $r$ for $q = -0.3, \; -0.5, \; -0.64, \; \text{and} \; -0.74$. 
 In the four panels, each graph represents the eigenvalue plotted in a specific plane defined by a constant value of $\theta$. In particular, the dotted line refers to the eigenvalue plotted in the plane $\pi/2$. }
\label{fig:TheEigenvaluesLambdaII}
\end{figure}

In plots \ref{fig:ThreeEigenvaluesA} and \ref{fig:TheEigenvaluesLambdaII}, we have followed the procedure described above to find regions of repulsive gravity. In Figure \ref{fig:ThreeEigenvaluesA} we plotted the eigenvalues for different values of the deformation parameter $q$: $-0.30$, $-0.50$, $-0.64$, and $-0.74$. In all these cases, we note that eigenvalues vanish at infinity. Moreover, as we approach the central source, the eigenvalues will increase very rapidly until they become infinity at the singularity, indicating a pure attractive behavior of gravity. Similarly, we plotted the eigenvalue $\Lambda_{S}$ corresponding to the Schwarzschild space-time. It has the same features as the last ones. Conversely, the second eigenvalues $\Lambda_{II}$ shows the presence of a zone with repulsive gravity. 

 In figure \ref{fig:TheEigenvaluesLambdaII}, we plotted the eigenvalue $\Lambda_{II}$ of the $q$-metric Riemann tensor for different values of $q$ as a function of the radial coordinate $r$, for different values of the angular coordinate $\theta$. We see that for $q$ fixed, the slope of the eigenvalue of the curvature tensor changes with the radial coordinate. We also observe that the extreme value of the curves moves toward the center of the coordinates as the values of the angular coordinate decreases from $\pi/2$ to 0.
 
 In the particular case, when $\theta=\pi/2$ and $q=-0.30$, we note that as we approximate from infinity towards the origin of coordinates, the slope exchanges its sign at $r = 2.415914727$, indicating the character of gravity has changed. Furthermore, this eigenvalue vanishes in $ r=2.145298686$, which implies that repulsive and attractive forces balance at such a point. 
 
A similar behavior of the eigenvalue is observed in Figures \ref{fig:TheEigenvaluesLambdaII}(b) - \ref{fig:TheEigenvaluesLambdaII}(d) for the values of the deformation parameter given by  $q=-0.50$, $q=-0.64$ and $q=-0.74$ . In the case in which $q=-0.50$ and $\theta = \pi/2$ as we approximate from infinity towards the origin of coordinates, we see that the slope exchanges their sign at $r = 2.412590872$, indicating the character of gravity has changed. Additionally, this eigenvalue vanishes in $ r= 2.151387819$, which implies that repulsive and attractive forces balance at such a point. The eigenvalue $\Lambda_{II}$ shows a similar behavior when $q=-0.64$ or $q=-0.74$. In these cases, the repulsion radii are $r = 2.363197880$ and $r = 2.307053260$, respectively. Whereas in these cases, repulsive and attractive forces balance each other at points $ r= 2.130157882$ and $ r= 2.104320276$, respectively. It is straightforward to note that the repulsion radius moves to the center of the coordinates as the values of $q$ decrease negatively. 
 
\begin{figure}[h!]
\centering
\includegraphics[width = 0.9\textwidth]{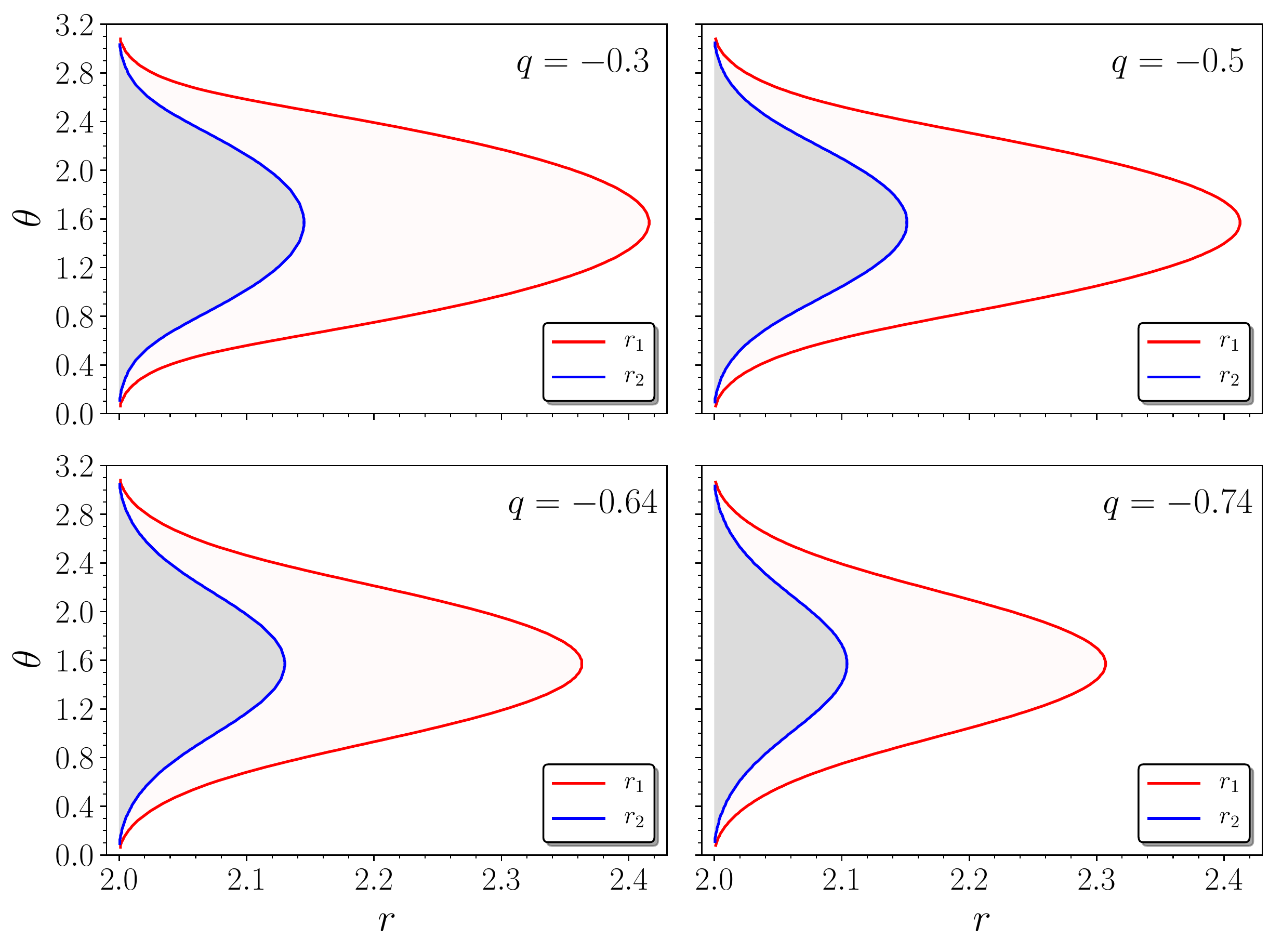}
\caption{plots for all $\theta$ values as a function of $r$ coordinate where the repulsive gravitational effects take place, for $q = -0.3, \; -0.5, \; -0.64, \; \text{and} \; -0.74$. In the four panels the grey region, delimited by the naked singularity at $r = 2M$ and $r = r_{2}$, corresponds to the zone where the eigenvalue exchanges its sign; the red zone, delimited by $r = r_{2}$ and $r = r_{1}$, corresponds to the region where the slope of the eigenvalue exchanges its signs. The white region beyond $r_1$ corresponds to the zone where gravity is completely attractive.  \label{fg:rep_zone}}
\end{figure}

With the aim of appreciating with more detail regions where repulsive gravitational effects take place, in figure \ref{fg:rep_zone} we show all values for $\theta$ as a function of $r$ where the eigenvalue and its slope exchange their signs for $q = -0.3,\; -0.5,\; -0.64 \; \text{and} \; -0.74$. In these plots we show two regions: the first one (grey zone), is delimited by $r = 2M$ and $r = r_{2}$ which correspond, respectively, with outer naked singularity and the radii where the eigenvalue vanishes; the second one (red zone), is delimited by $r = r_{2}$ and $r = r_{1}$, where the last one represents the radii where the slope exchange its sign. Moreover, for $r > r_1$ (white zone) gravity has a purely attractive behavior. 
\\ \\ \\
From figure \ref{fg:rep_zone} we can observe that for all values of $q$ employed, the maximum radius of the grey and red zone is located at $\theta = \pi / 2$, while the minimum occurs near the poles, i.e., $\theta = 0$ and $\theta = \pi$; besides, we can notice that as $q$ is reduced both zones also decrease. Moreover, we can see that the naked singularity will always be surrounded by a region with repulsive effects. Based on these results, and the analysis of the trajectories of the photons that we will carry out in the next section, we will determine whether or not the effects of repulsive gravity can be observed in the shadow produced by the naked singularity.
\section{Photon motion equations and shadow formulation}
\label{sec:PME}

The study of null geodesics is important to predict, understand, and interpret observational results. In particular, null geodesics around black holes and other exotic compact objects are of special interest, because the path followed by the photons could modify the optical perception of a distant observer. For this reason, we want to determinate the motion equations of test particles moving in a curved space-time. They can be obtained from the Hamilton equations, which read as
\begin{equation}
\dot{p}_{\alpha} = - \frac{\partial \mathscr{H}}{\partial x^{\alpha}}, ~~~~~ \dot{x}^\alpha = \frac{\partial \mathscr{H}}{\partial p_{\alpha}}, \label{eq:HamEq}
\end{equation}
where $p_\alpha$ and $x^\alpha$ are the components of the four-momentum and the four-position, respectively, and the overdot means derivate respect to an affine parameter. Additionally, we have defined the Hamiltonian in the standard way 
\begin{equation}
\mathscr{H}(x^{\alpha},p_{\alpha}) = p_{\alpha}\dot{x}^{\alpha} - \mathscr{L}(x^{\alpha},\dot{x}^{\alpha})= \frac{1}{2} g^{\alpha\beta} p_\alpha p_\beta = -\frac{1}{2}\mu^2, \label{eq:Ham}
\end{equation}
being $g^{\alpha\beta}$ the contravariants components of the metric tensor and $\mu$ the rest mass of the test particles. It is worth mentioning that $\mu=1$ and $\mu=0$ correspond, respectively, to time-like and null-like particles. On the other hand, the space-time we are studying is static and axially symmetric, which implies that there are two conserved quantities: the energy and the azimuthal component of the angular momentum. Each one is given by $p_t = -E$ and $p_\phi = L$, and are related with two cyclic coordinates, $t$ and $\phi$, respectively. It is important to consider that there is not a third integral of motion due to the deformation parameter, $q$, in \eqref{eq:q-metric}.

Now, to obtain the apparent shape of a compact object described by the $q$-metric, we define $\left \lbrace \hat{\varepsilon}_\alpha \right \rbrace$ as an orthogonal coordinate basis in a static and axisymmetric space-time, such that the inner product obeys 
\begin{gather}
\mathbf{g}(\hat{\varepsilon}_\alpha,\hat{\varepsilon}_\beta) = \hat{\varepsilon}_\alpha\cdot\hat{\varepsilon}_\beta = g_{\alpha\beta},
\end{gather}
where $g_{\alpha\beta}$ are the covariants components of the metric tensor, and $\hat{\varepsilon}_\alpha\cdot\hat{\varepsilon}_\beta = 0$ when $\alpha\neq\beta$. Moreover, we can find a relation with a locally inertial frame through a transformation matrix
\begin{gather}
\hat{e}_\alpha = \Lambda_\alpha^{\,\,\beta}\hat{\varepsilon}_\beta, 
\end{gather}
where $\left \lbrace \hat{e}_\alpha \right \rbrace$ satisfy the Lorentzian inner product
\begin{gather}
{\mathbf{\eta}}(\hat{e}_\alpha,\hat{e}_\beta) = \hat{e}_\alpha\cdot\hat{e}_\beta = \eta_{\alpha\beta},
\end{gather}
being $\eta_{\alpha\beta}$ the components of the Minkowski metric. Since $\left \lbrace \hat{\varepsilon}_\alpha \right \rbrace$ is orthogonal, any collinear vector will be orthogonal to each other. For that reason, we defined the matrix as
\begin{gather}
\left[\Lambda_{\alpha}^{\,\,\beta}\right]=\left(\begin{array}{cccc}{A_{t}} & {0} & {0} & {0} \\ {0} & {B_{r}} & {0} & {0} \\ {0} & {0} & {C_{\theta}} & {0} \\ {0} & {0} & {0} & {D_{\phi}}\end{array}\right),
\end{gather}
where each element is a function of the metric tensor components. Using the Minkowski normalization, we obtain
\begin{gather}
A_t = \frac{1}{\sqrt{-g_{tt}}}, \quad B_r = \frac{1}{\sqrt{g_{rr}}}, \quad = C_\theta = \frac{1}{\sqrt{g_{\theta\theta}}}, \quad D_{\phi} = \frac{1}{\sqrt{g_{\phi\phi}}}. \label{eq:components}
\end{gather}
The next step is sending the locally inertial observer far away from the gravitational source, and relating the components of the four-moment $\mathbfcal{P}$ in this frame with the components of the four-moment associated to the basis $\left\lbrace \hat{\varepsilon}_\alpha \right\rbrace$ as follows	
\begin{gather}
\mathcal{P}^{\alpha}=\eta^{\alpha \mu} \Lambda_{\mu}^{\,\,\beta} p_{\beta}. \label{eq:momenta}
\end{gather}
Now, proceeding as in \cite{cunha}, we can obtain a general expression for all points ($x,y$) in the observer's plane
\begin{gather}
x = -r_0\frac{\mathcal{P}^\phi}{\mathcal{P}^r}, \qquad y = r_0\frac{\mathcal{P}^\theta}{\mathcal{P}^r}. \label{eq:xy}
\end{gather}
Besides, it is easy to show that $\mathcal{P}^r = \mathcal{P}^t$ when the observer is away from the source, since in this frame $\mathcal{P}^r = | \vec{\mathcal{P}}|$, which corresponds to the spatial part of the four-momentum, and due to hamiltonian constraint, $\mathscr{H} = 0$, we have
\begin{gather}
\frac{1}{2}|\mathbfcal{P}| = 0 \quad \Rightarrow \quad \mathcal{P}^t = \sqrt{(\mathcal{P}^r)^2 + (\mathcal{P}^\theta)^2 + (\mathcal{P}^\phi)^2} = |\vec{\mathcal{P}}|. \label{eq:ptpr}
\end{gather}
Once the range of observation is defined, we integrate the motion equations in \eqref{eq:HamEq} with initial coordinates $(t_0,\,r_0,\,\theta_0,\,\phi_0)$ and initial momentums
\begin{gather}
\begin{aligned} p_{\theta} &=\frac{y \sqrt{g_{\theta \theta}}}{r_{0}}, 
& \qquad
E &=\sqrt{-g_{tt}}+\frac{x g_{t \phi}}{r_{0} \sqrt{g_{\phi \phi}}}, 
\\ \\ 
L &=-\frac{x \sqrt{g_{\phi \phi}}}{r_{0}}, 
& & 
p_{r}=\sqrt{g_{r r}\left[1-\left(\mathcal{P}^{\theta}\right)^{2}-\left(\mathcal{P}^{\phi}\right)^{2}\right]},
\end{aligned}
\end{gather}
which were obtained combining \eqref{eq:components}-\eqref{eq:ptpr}. Here, we assumed $\mathcal{P}^t = \mathcal{E} = 1$ because it is related to the frequency of the photon, and the gravitational red-shift effects are ignored. The algorithm used to obtain the apparent shape of the compact object is based on the ``backward ray-tracing'' method, where the observer is located at the origin of null geodesics, and the motion equations are integrated backward in time using a Runge-Kutta time integrator with Dormand-Prince coefficients and adaptive step. The equations are integrated until the photons reach the naked singularity, numerically it occurs when $r \leq 2m + \delta r$, where $\delta r$ is a numerical buffer. Besides, the accuracy of the method along the integration process is guaranteed by the Hamiltonian constraint. Finally, in the code we set the value $m=1$ for all the simulations, because this parameter only rescale the shadow size.

\section{$q$-metric shadow}
\label{sec:QM}
Most of the information that we can get from outer space is obtained from electromagnetic radiation; therefore, the study of null geodesics is of great astrophysical interest. In the case of black holes, due to their nature, it is not possible to obtain a direct image of them, because they do not radiate. Consequently, the information that can be obtained from these objects comes from the photons that orbit close to the event horizon and escape from the gravitational attraction. The trapped radiation forms a black spot on the bright background that illuminates the vision of a distant observer, whose telescopes are pointed to the black hole. This dark spot is called the shadow of a black hole and is the closest to an image that we can get from them. In this way, following \cite{Bohn_2015}, we assumed a celestial sphere as a bright source that englobes the observer and is concentric with the compact object. This sphere is sectioned in four quadrants with a specific color: blue, green, yellow, and red, and the notion of curvature is provided by a black mesh with latitude and longitude constant lines. Additionally, each point ($x,y$) on the image plane, which corresponds uniquely with just one set of initial conditions, is painted following the section of the sphere where the photon strikes. Moreover, the local inertial observer is located on the equatorial plane of the celestial sphere at $r = 100 M$, and the black spot in the middle corresponds to the shadow of the compact object. 

\begin{figure}[ht]
\centering
\includegraphics[width = 0.9\textwidth]{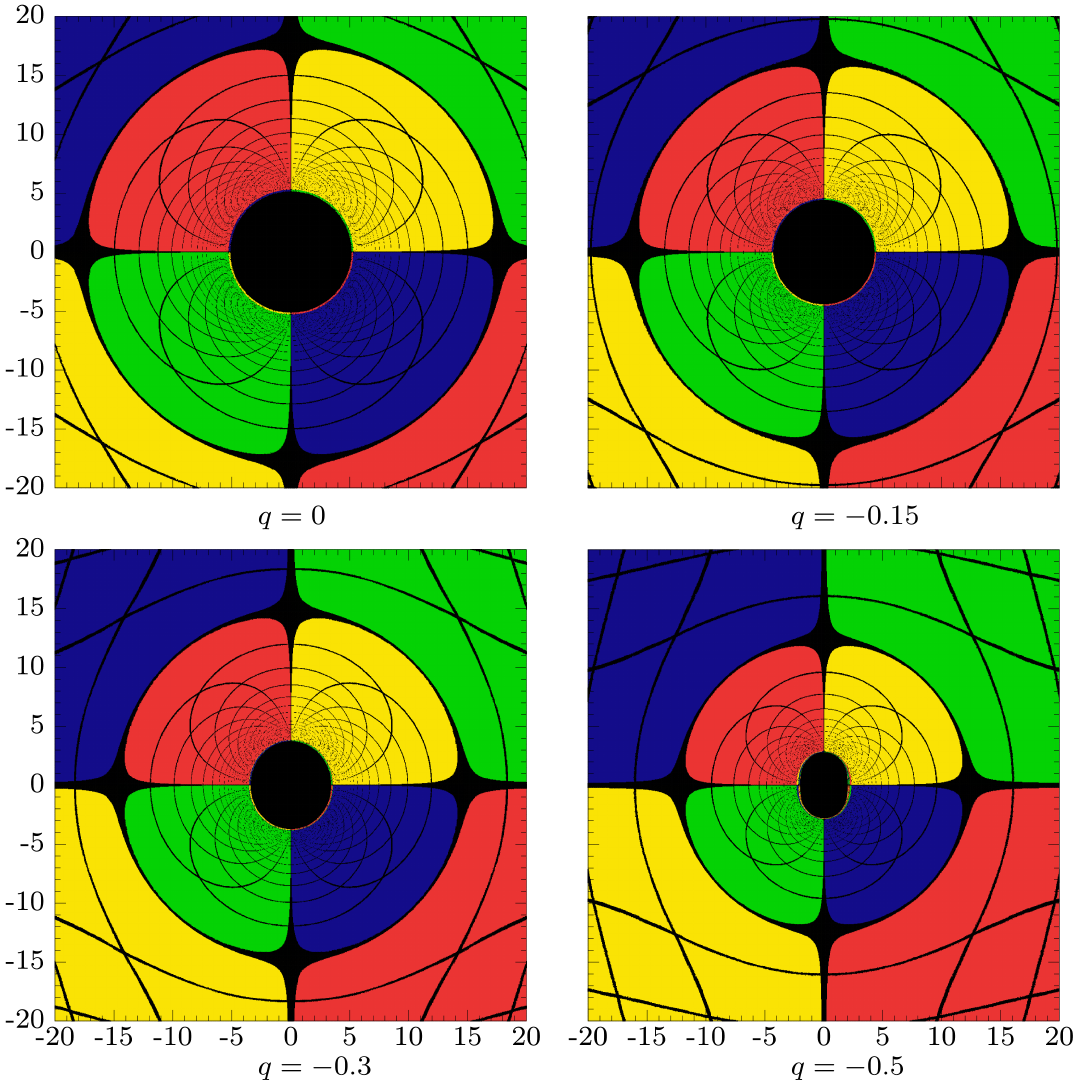}
\caption{shadow and gravitational lensing generated by a compact object described by the $q$-metric in the domain of integration $ x,y \in [ -20, 20 ]$ for $0 \geq q \geq -0$.5. The top-left panel shows the Schwarzschild space-time for $q = 0$, the top-right panel shows the change of the Einstein ring for $q = -0$.15, and the bottom row shows how the shape of the shadow is getting more prolate as $q$ decreases. In the four panels, the trajectories of the photons are almost the same; nevertheless, in $q = -0$.5 there are new pseudo-rings on the border of the shadow. \label{fg:ql}}
\end{figure}

\begin{figure}[ht]
\centering
\includegraphics[width = 1\textwidth]{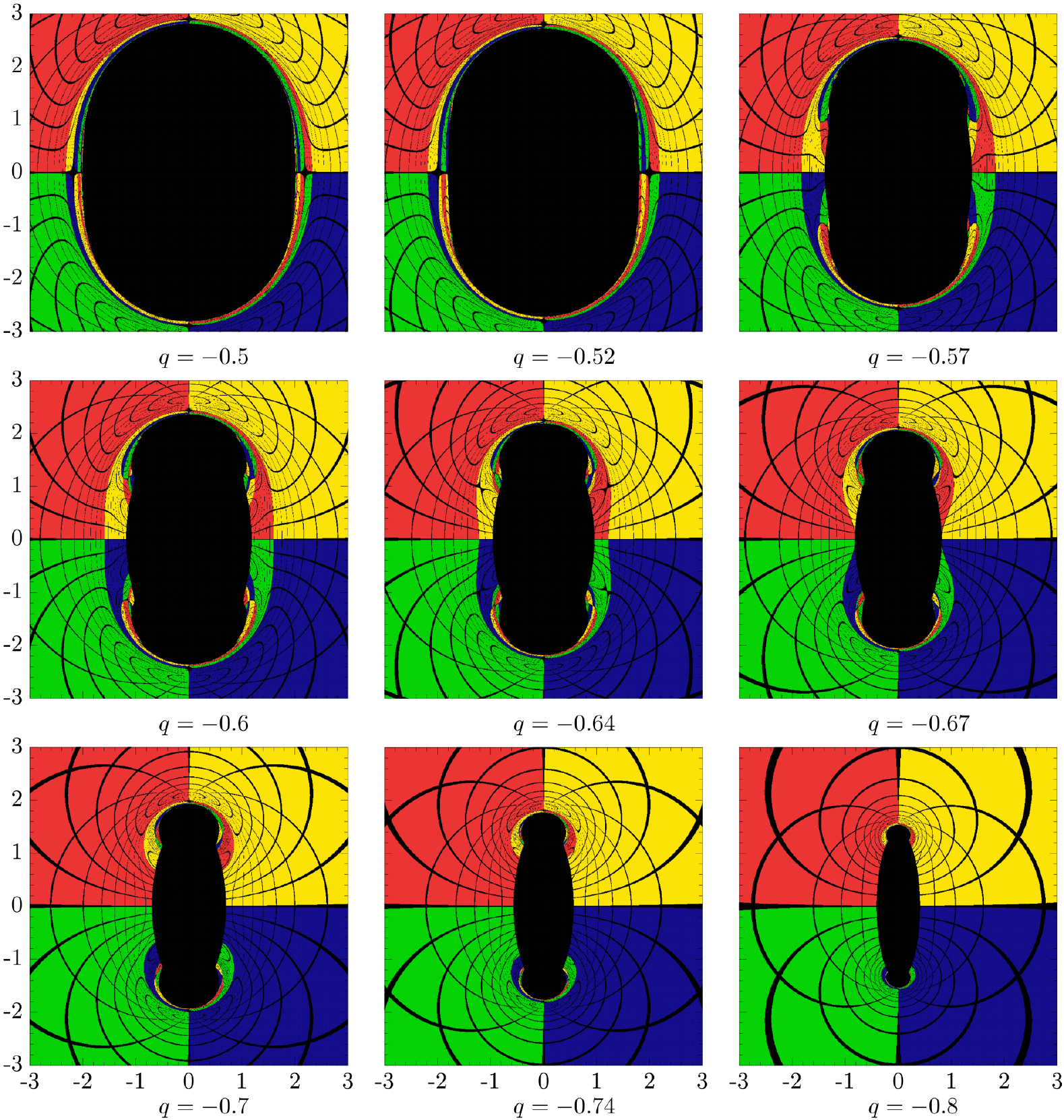}
\caption{shadow and gravitational lensing generated by a compact object described by the $q$-metric in the domain of integration $ x,y \in [ -3, 3 ]$ for $ -0.5$ $\leq q \leq -0.8$. The generation of pseudo-rings can be appreciated from $q = -0.5$ to $q = -0.6$. For $q = -0.64$ and $q = -0.67$ a waist is formed at $y = 0$, while for $q = -0.7$ the waist is broken. On the other hand, from $q = -0.57$ we can see the appearance of bulges at the upper and lower region of the shadow. Finally, the bulges and the shadow tend to disappear when $q$ decreases. \label{fg:qm}}
\end{figure}

\begin{figure}[h!]
\centering
\includegraphics[width = 1\textwidth]{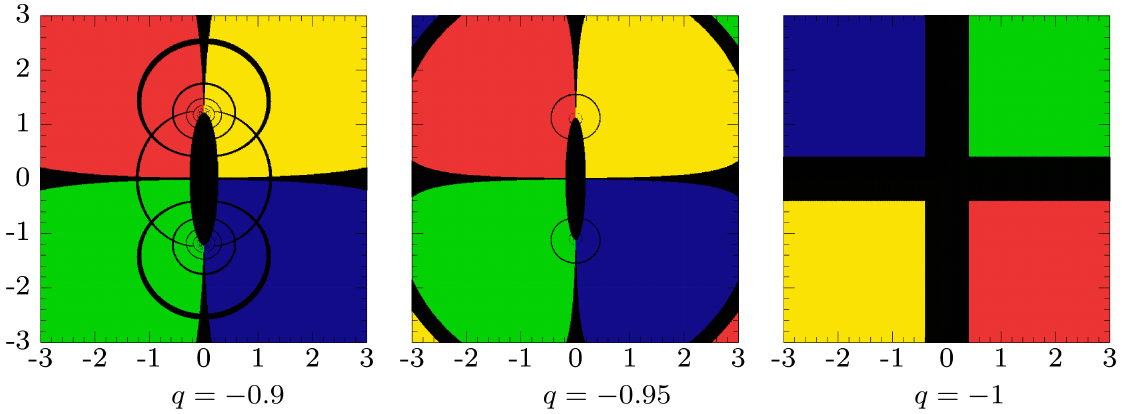}
\caption{shadow and gravitational lensing generated by a compact object described by the $q$-metric in the domain of integration $ x,y \in [ -3, 3 ]$ for $ -0.9$ $\leq q \leq -1$. For $q = -0.9$ the bulges of the shadow are reduced to two points from which the mesh is curved. For $q = -0.95$ the Einstein ring appears again. Finally, for $q = -1$ the $q$-metric reduces to Minkowski space-time. \label{fg:qs}}
\end{figure}

\begin{figure}[h!]
\centering
\subfloat{
  \includegraphics[width = 0.5\textwidth]{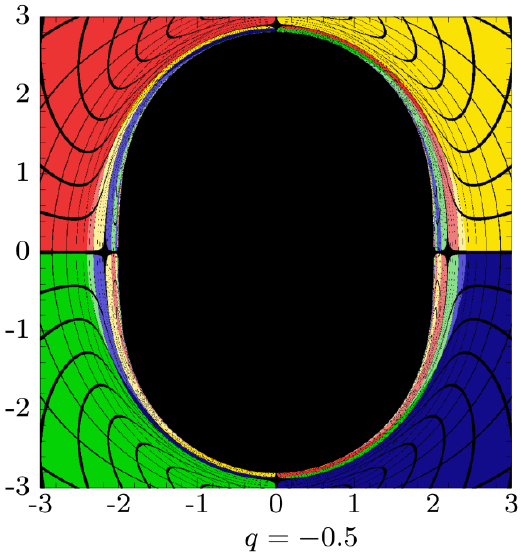}
}
\hspace{0mm}
\subfloat{
  \includegraphics[width = 0.5\textwidth]{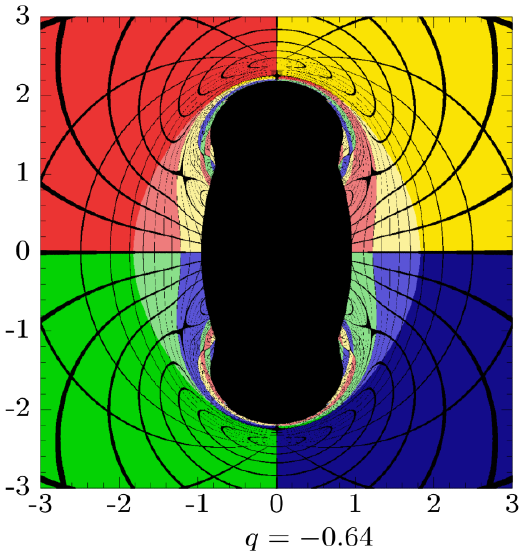}
}
\subfloat{
  \includegraphics[width = 0.5\textwidth]{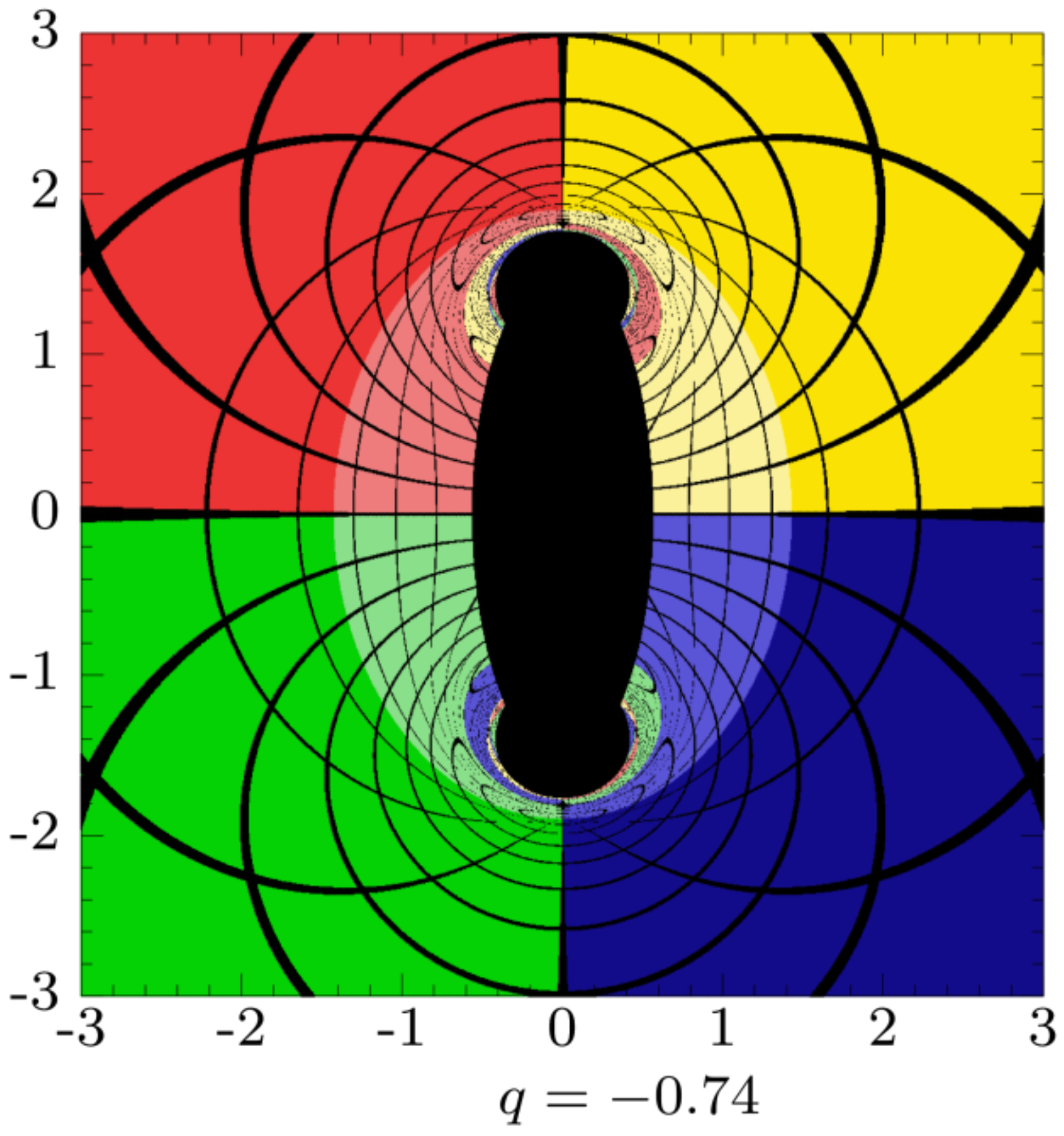}
}
\caption{shadow and gravitational lensing with repulsive zone for $ q = -0.5 $ (top), $ q = -0.64 $ (bottom left) and $ q = -0.74 $ (bottom right). Photons that orbit in a repulsive gravity region delimited by $ r_1 $ and $ r_2 $, and scape from the gravitational attraction, are painted in faded colors. For all values of $q$, the pseudo-rings and the shadow are inside of this region. Moreover, for $ q = -0.64 $ and $ q = -0.74 $, the bulges where the dependence of the system to initial conditions is strong are in this region as well. On other hand, photons that orbit in a repulsive zone between $r = 2m$ and $r_1$ fall to the naked singularity.}
\label{fig:repulsive_shadow}
\end{figure}

On the other hand, in order to appreciate the impact of the deformation on null geodesics, we set two numerical integration domains for different values of $q$
\begin{enumerate}
\item[I.] $ x,y \in [ -20, 20 ]$ for $0 \geq q \geq -0$.5,
\item[II.] $ x,y \in [ -3, 3 ]$ for $ -0$.5 $\leq q \leq -1$,
\end{enumerate}
and after solving the equations presented in section 3, we show in figures \ref{fg:ql}-\ref{fig:repulsive_shadow} the shadow and the gravitational lensing generated by a compact object, whose gravitational field is described by the $q$-metric.
In figure \ref{fg:ql}, corresponding to the domain I, since the $q$-metric reduces to the Schwarzschild space-time, the spherical symmetry is recovered for the case $q = 0$; however, negative values of $q$ make the shadow becomes prolate, and its size is reduced. This behavior is clear in the transition from $q = -0$.15 to $q =-0$.5. Also, when $q$ is reduced, the Einstein ring size decreases, but its shape does not change, and the gravitational lensing is not very different from the one in the Schwarzschild space-time case.

Now, let us consider the domain of integration II. As we can see in figure \ref{fg:qm}, for all values of $q$ in this range, we are observing inside the Einstein ring. There are very close points in the gravitational lensing that correspond to different colors, which implies a strong dependence of the system to the initial conditions: small variations of them change the trajectory of the photons significantly. In the top row of figure \ref{fg:qm}, it is clear that $q = - 0$.5 is a critical value since appears pseudo-rings near the edge of the shadow. When $q = -0$.57, these pseudo-rings are deformed, then are grouping in the bulges that appear at the upper and lower regions of the shadow. Additionally, the shadow is smaller, and its shape is deformed. For $q = -0$.64 the outer pseudo-ring forms a waist at $y = 0$. We can see that when $q = -0$.7, the waist is broken, and it is grouped in the bulges, while the size of the shadow decreases and becomes thinner. The behavior is quite similar for lower values of $q$. Furthermore, in figure \eqref{fg:qs}, we can see that for $q = -0$.9, the bulges are reduced to two points from which the mesh is curved. In the case $q = -0$.95, we can appreciate how the Einstein ring appears again, because as we said before, it is getting smaller, and when $q = -1$, we obtain the flat space-time. Besides, the cases in figures \eqref{fg:ql}, \eqref{fg:qm}, \eqref{fg:qs} coincide with a rod-like source for $-1 < q < 0$, according to \cite{kodama2003global}. 
\\ \\
\\
On the other hand, it is necessary to mention that if the repulsive gravitational effects could be measured, its influence over null geodesics would be observed through the gravitational lensing. For this reason, it is important to distinguish the regions in the observer's sky where the repulsive effects take place. To achieve it, from the results presented in section \ref{sec:ST}, in figure \ref{fig:repulsive_shadow} we show the regions in the gravitational lensing where the repulsive gravity appears for $q = -0.5, \; -0.64, \; \text{and} \; -0.74 $, following the regions shown in figure \ref{fg:rep_zone}.
\\ \\ 
In figure \ref{fig:repulsive_shadow}, the repulsive gravity zone is painted in fade colors (corresponding to the red zone in figure \ref{fg:rep_zone}), from which it is important to emphasize that the pseudo-rings, the waist and the regions where the strong dependence of the system to the initial conditions are related to the repulsive effects. Further, from our simulations, we find that the naked singularity inevitably traps null geodesics in the repulsive zone where the eigenvalue exchange its sing (grey zone in figure \ref{fg:rep_zone}). In consequence, it would be impossible to obtain information about this region. It is worth mentioning that a meticulous inspection of the Riemann tensor's eigenvalues allows us to find repulsive effects zones for all values of $q$; however, for $q > -0.5$, the naked singularity traps all photons that orbit in repulsive zones. Hence, from an observational point of view, these values are useless to distinguish the signature of repulsive gravity produced by a naked singularity. 
\newpage
Finally, we only carry out a characterization of the impact of deformation parameter on the shadow described by the $q$-metric for negative values of $q$, since oblate deformations do not represent significant differences respect to the Schwarzschild case. Besides, the prolate deformation influence is appreciable only near the gravitational source for $q<-0.5$, unlike other exotic solutions, where slight deformations of the source represent a significant change in the shadow shape \cite{wang}. 

\section{Conclusion}
\label{sec:conclusions}

In this work, we studied the gravitational field of a source with an arbitrary mass deformation parameter $q$. This space-time is described by the $q$-metric, which is the most straightforward generalization of the Schwarzschild metric containing naked singularities. 

First of all, we calculated the eigenvalues of the Riemann tensor in the representation $SO(3, C)$ to determine, in an invariant way, the regions where the gravitational repulsive effects of the compact object take place, for  $q = -0.3, \, -0.5, \, -0.64, \, \text{and} \, -0.74$. The existence of an extremum in the eigenvalue $\Lambda _{II}$ predicts the presence of repulsive gravity; moreover, the vanishing of $\Lambda _{II}$ indicates the point at which attractive gravity becomes entirely compensated by the action of repulsive gravity. From this analysis, we found that near to the source there is a region with repulsive gravity, and the repulsion radius moves to the origin of coordinates as $q$ decreases.

Subsequently, we numerically solved the motion equations for null geodesics to analyze the shadow, the gravitational lensing, and the Einstein ring produced by the $q$-metric. We found that as the $q$ parameter decreases from $q = 0$ to $q=-0.5$, the shadow exhibits a prolate deformation. The Einstein ring decreases without changing its shape. The gravitational lensing is not much different from the one in the Schwarzschild space-time case. On the other hand, we noticed that $q=-0.5$ behaves as a critical value because the appearance of pseudo-rings' close to the edge of the shadow. Moreover, for even smaller values of $q$, we observed that the innermost pseudo-ring misshapen until a waist is formed, which breaks forming two bulges at the upper and lower sides of the shadow where there is a strong dependence of the system to the initial conditions. Besides, the size of the shadow becomes thinner as the $q$ parameter gets smaller, and for $q = -1$, the Minkowski space-time is recovered.  

According to the above and from our simulations, the pseudo-rings, the waist, and the bulges on the gravitational lensing could be evidence of the repulsive gravity effects, which only correspond to the regions where the slope of eigenvalue $\Lambda _{II}$ exchanges its sign. Furthermore, we found that photons that orbit in zones where $\Lambda _{II} < 0$ are trapped by the naked singularity; in other words, this zone constitutes the edge of the shadow for these values of $q$. Otherwise, for all values of $q > -0.5$ repulsing zones appear as well; however, it is impossible to obtain information of any repulsive effects due to all photons in these regions also are trapped by the naked singularity. 

Finally, the crucial question left unanswered by our study, and that opens a study window, is how these highly dynamical effects, close up of shadow silhouette, are related with the repulsive effects produced by the naked singularity. Answering this will require more understanding of the repulsive effects presented in naked singularities.

\section*{Acknowledgments} J.A.A-V and J.M.V-C, want to thanks the financial support from Universidad Industrial de Santander. O. M. P. wants to thanks the financial support from COLCIENCIAS under the program Becas Doctorados Nacionales 647 and Universidad Industrial de Santander. F.D.L-C acknowledges support from Vicerrector\'ia de Investigaci\'on y Extensi\'on - Universidad Industrial de Santander, under Grant No. 2493 and from COLCIENCIAS, Colombia, under Grant No. 8863. A.C.G-P. is thankful to the Departamento de Gravitaci\'on y Teor\'ia de Campos (ICN-UNAM) for its hospitality during his research fellowship. Also, A.C.G-P would like to Hernando Quevedo for useful comments and discussions. This work was partially supported by Programa Capital Semilla para Investigaci\'on, Proyecto 2490, VIE-UIS .

\section*{References}

\providecommand{\newblock}{}

\end{document}